\documentclass[conference,a4paper]{IEEEtran}
\usepackage{cite}
\ifCLASSINFOpdf
   \usepackage[pdftex]{graphicx}
  \graphicspath{{../pdf/}{../jpeg/}}
   \DeclareGraphicsExtensions{.pdf,.jpeg,.png}
\else
   \usepackage[dvips]{graphicx}
   \graphicspath{{../eps/}}
   \DeclareGraphicsExtensions{.eps}
\fi
\usepackage[cmex10]{amsmath}

\usepackage{caption}
\usepackage[caption=false,font=footnotesize]{subfig}
\usepackage{fixltx2e}
\usepackage{stfloats}
\begin{document}
\title{Application of computational breast phantoms to evaluate reconstruction methods for fluorescence molecular tomography}

\author{\IEEEauthorblockN{
Yansong Zhu\IEEEauthorrefmark{1},   % 1st author, 1st affiliations
Abhinav K. Jha\IEEEauthorrefmark{2},\IEEEauthorrefmark{4} and   % 2nd author, 2nd affiliations
Arman Rahmim\IEEEauthorrefmark{1}\IEEEauthorrefmark{2}    % 3rd author, 2nd affiliations
}                                     % ...
\\
\IEEEauthorblockA{\IEEEauthorrefmark{1}% 1st affiliations
Department of Electrical and Computer Engineering, 
Johns Hopkins University, Baltimore, USA}
\IEEEauthorblockA{\IEEEauthorrefmark{2}% 2nd affiliations
Department of Radiology, 
Johns Hopkins University, Baltimore, USA}
\IEEEauthorblockA{\IEEEauthorrefmark{4}% 4th affiliation
Corresponding author: ajha4@jhmi.edu}
}


%Telephone: (800) 555--1212, Fax: (888) 555--1212}


\maketitle

\begin{abstract}
%\boldmath
Fluorescence molecular tomography (FMT) has potential of providing high contrast images for breast tumor detection. Computational phantom provides a convenient way to a wide variety of fluorophore distribution configurations in patients and perform comprehensive evaluation of the imaging systems and methods for FMT. In this study, a digital breast phantom was used to compare the performance of a novel sparsity-based reconstruction method and Tikhonov regularization method for resolving tumors with different amount of separation. The results showed that the proposed sparse reconstruction method yielded better performance.
This simulation-based approach with computational phantoms enabled an evaluation of the reconstruction methods for FMT for breast-cancer detection. 
\end{abstract}

%{\smallskip \keywords component; formatting; style; styling; insert}

% \fi
\IEEEpeerreviewmaketitle

\section{Introduction}
Fluorescence imaging is an emerging optical technique that provides high contrast image with the use of fluorophores. Previously, Indocyanine Green (ICG) has been reported for fluorescence image-guided surgery for breast cancer\cite{JSO:JSO21943}. Tomography enables 3D visualization and estimation of fluorescence distribution. Recently, clinical studies have been conducted to study the feasibility of fluorescence molecular tomography (FMT) for breast tumor detection\cite{corlu2007three}. The advent of sparse reconstruction methods have shown potential of providing more accurate estimation of fluorescence distribution and intensity compared to conventional method in FMT\cite{dutta2012joint}. However, studies are required to assess whether the sparse reconstruction methods indeed yield improved performance for FMT-based breast-cancer imaging. Computational phantom provides a comparatively easier way to conduct these studies by allowing modeling of different fluroscence distributions in breast-cancer populations as opposed to having to actually image patients. In this study, we conducted a computational human-phantom-based study to compare the performance of sparse reconstruction method and a traditional Tikhonov regularization method for resolving tumors in patient images.

\section{Theory}
\subsection{Forward model}
The forward model of FMT can be described with two equations representing light propagation from source to fluorophore and from fluorophore to detector, respectively. Let $\mathbf{r_s}$, $\mathbf{r}$ and $\mathbf{r_d}$ denote positions of source, fluorophore and detector respectively, $g_{ex}(\mathbf{r_s},\mathbf{r})$, $g_{em}(\mathbf{r},\mathbf{r_d})$ denote the Green's function of excitation and emission light, $x(\mathbf{r})$ denote fluorescence yield, $s(\mathbf{r_s})$ denote source function, $\Omega_s$, $\Omega$ denote the support of source and fluorophore positions, respectively, and $\phi_x(\mathbf{r})$, $\phi_m(\mathbf{r})$ denote fluence of source and fluorophore, then the forward model is written as:
\begin{equation}
\phi_x(\mathbf{r})=\int_{\Omega_s}{g_{\mathrm{ex}}(\mathbf{r_s},\mathbf{r})s(\mathbf{r_s})d^3r_s},
\label{equ:fwd1}
\end{equation}
and
\begin{equation}
\phi_m(\mathbf{r_d})=\int_{\Omega}{g_{\mathrm{em}}(\mathbf{r},\mathbf{r_d})x(\mathbf{r})\phi_x(\mathbf{r})d^3r}.
\label{equ:fwd2}
\end{equation}
From equation (\ref{equ:fwd1}) and (\ref{equ:fwd2}) we can derive the following linear matrix equation for measurements and unknown fluorophore distribution:
\begin{equation}
\mathbf{\Phi}=\mathbf{Gx},
\label{equ:fwdmatrix}
\end{equation}
where $\mathbf{G}$ is the system matrix.
%\begin{equation*}
%\mathbf{\Phi}=
%\begin{bmatrix}  
%\phi^1_m\\ 
%\phi^2_m\\
%\vdots \\ 
%\phi^M_m
%\end{bmatrix},
%\end{equation*}
%\begin{equation*}
%\mathbf{G}=
%\begin{bmatrix} 
%g^{11}_{\mathrm{em}}\phi_x^1&\dots &g^{1N}_{\mathrm{em}}\phi_x^1\\
%\vdots & \vdots& \vdots\\
%g^{M1}_{\mathrm{em}}\phi_x^1 & \dots & g^{MN}_{\mathrm{em}}\phi_x^1\\
%g^{11}_{\mathrm{em}}\phi_x^2&\dots &g^{1N}_{\mathrm{em}}\phi_x^2\\
%\vdots & \vdots &\vdots\\
%g^{M1}_{\mathrm{em}}\phi_x^{N_s} & \dots & g^{MN}_{\mathrm{em}}\phi_x^{N_s}\\
%\end{bmatrix},
%\end{equation*}
%and
%\begin{equation*}
%\mathbf{x}=
%\begin{bmatrix}  
%\eta\mu^1_{af}\\ 
%\eta\mu^2_{af}\\
%\vdots \\ 
%\eta\mu^N_{af}
%\end{bmatrix}.
%\end{equation*}
\subsection{Proposed sparsity-driven reconstruction method}
The inverse problem of FMT is usually ill-posed and regularization is required in order to solve the fluorescence distribution. Tikhonov regularization method is widely used and it can be expressed as follows:
\begin{equation}
\hat{\boldsymbol{x}}=\rm{arg}\min_x\{\|\mathbf{Gx}-\mathbf{\Phi}\|^2_2+\|\Gamma\mathbf{x}\|^2\},
\label{equ:recfunc}
\end{equation}
where $\Gamma$ is typically chosen as $\lambda\mathit{I}$. $\mathit{I}$ is identity matrix.\\
\indent Breast tumor tends to concentrate in a small region compared with the whole breast, and fluorophore is designed to target the tumor. Thus, the inverse problem of FMT for breast tumor can also be regarded as a sparse reconstruction problem. We propose a sparse-reconstruction approach similar to another approach we developed for transcranial small-animal imaging previously\cite{Yansong17}.  The basic idea is to minimize the following $\ell_1$ regularized problem to solve for $\mathbf{x}$:
\begin{equation}
\hat{\boldsymbol{x}}=\rm{arg}\min_x\{\frac{1}{2}\|\mathbf{Gx}-\mathbf{\Phi}\|^2_2+\lambda\|\mathbf{x}\|_1\}.
\label{equ:recfunc}
\end{equation}
\indent Sparse reconstruction algorithm requires that measurement matrix $\mathbf{G}$ should be nearly orthogonal. For this purpose, the coherence of G needs to be reduced, for which we use the truncated SVD.
Fast iterative shrinkage-thresholding algorithm (FISTA) is then used to find sparse solution for equation (\ref{equ:recfunc}). Finally maximum-likelihood expectation maximization algorithm is implemented for noise reduction.
\section{Computational studies with breast phantoms}
Simulation-based experiments were conducted to study the ability of FMT for resolving breast tumor. A breast phantom generated from MR database was used in this experiment. The phantom was segmented into four different tissues: blood vessels, skin layer, fat, and fibroglandular tissues. Different optical properties were assigned to each type of tissue, as described in \cite{Lou17}. Two tumors of size 1.6 cm were inserted in the breast phantom at difference seperation and fluorophore was assumed to concentrate in the tumors. 40 laser sources and 40 detectors were positioned around the phantom. Monte Carlo method was implemented to calculate the forward model\cite{Fang:09}. Two different methods, Tikhonov regularization method and the proposed sparse reconstruction method were used for reconstruction of FMT.  
\begin{figure*}[!t]
\centerline{\subfloat[True image]{\includegraphics[width=1.8in]{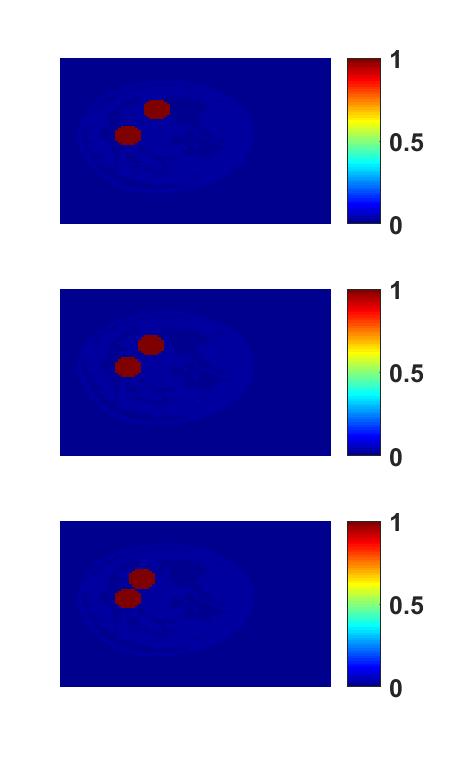}%
\label{fig_first_case}}
\hfil
\subfloat[Tikhonov regularized reconstruction]{\includegraphics[width=1.8in]{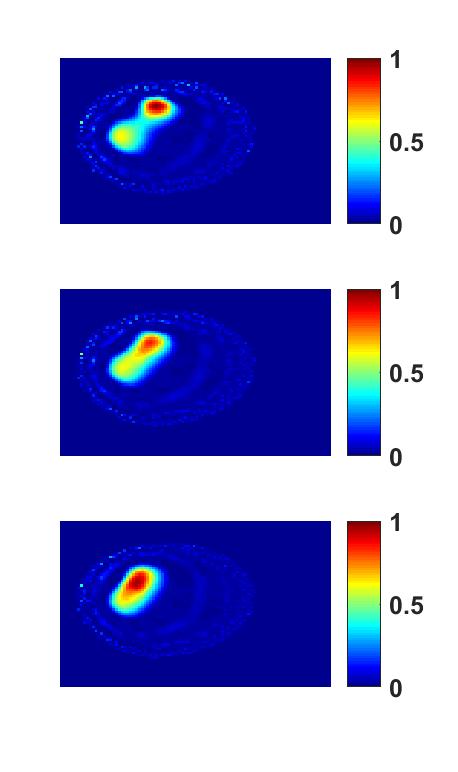}%
\label{fig_second_case}}
\hfil
\subfloat[Sparse reconstruction]{\includegraphics[width=1.8in]{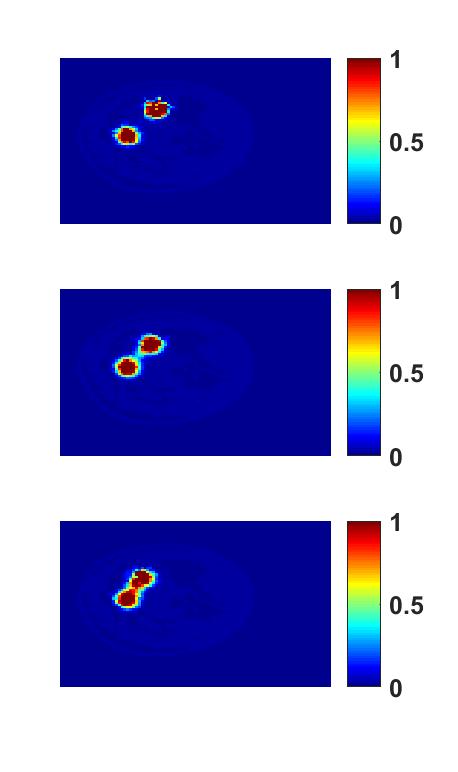}%
\label{fig_second_case}}}
\caption{Comparison of reconstruction with Tikhonov regularization method and the sparse reconstruction method. From top row to bottom row, the distances of the center of two tumors are 2.5 cm, 2 cm and 1.7 cm, respectively }
\label{fig_sim}
\end{figure*}
The result is shown in Figure (\ref{fig_sim}).
\section{Results and Conclusions}
From figure (\ref{fig_sim}), we notice that for larger separation between tumors (2.5cm), both Tikhonov method and sparse reconstruction method can differentiate the two tumors. When the separation between the tumors decreases, Tikhonov regularization method oversmooths the edge of the tumors, making it hard to distinguish them. However, the sparse reconstruction method is still able to resolve the two tumors. In addition, we also notice that compared to Tikhonov reconstruction method, sparse reconstruction provides more accurate estimation of fluorophore distribution and signal intensity.\\
\indent In conclusion, this simulation-based approach with computational phantoms enables an evaluation of the reconstruction methods for FMT for breast-cancer detection. In future, we propose to conduct objective evaluation studies to further evaluate the performance of these reconstruction methods\cite{jha2013ideal}.


\section*{Acknowledgment}
This work was supported by the NIH BRAIN Initiative Award R24 MH106083.



\bibliographystyle{IEEEtran}
\bibliography{BPbib}

\end{document}